\documentclass{webofc}
\usepackage[varg]{txfonts}   
\usepackage{color}

\newcommand {\snn}	{\sqrt{s_{_{\rm NN}}}}

\newcommand {\hijing}	{Hijing}
\newcommand {\pythia}     {Pythia}
\newcommand {\urqmd}     {UrQMD}
\newcommand {\ampt}	{AMPT}
\newcommand {\amptdef}	{AMPT-Def}
\newcommand {\amptsm}	{AMPT-SM}

\newcommand {\Zr}	{$^{96}$Zr}
\newcommand {\Ru}	{$^{96}$Ru}
\newcommand {\RuRu}	{$^{96}_{44}$Ru+$^{96}_{44}$Ru}
\newcommand {\ZrZr}	{$^{96}_{40}$Zr+$^{96}_{40}$Zr}
\newcommand {\RuZr}	{Ru+Ru/Zr+Zr}
\newcommand {\Pb}	{$^{208}$Pb}
\newcommand {\rnp}	{$\Delta r_{\rm np}$}
\newcommand {\Lc}	{L(\rho_{c})}

\newcommand {\Nch}	{N_{\rm ch}}
\newcommand {\pt}	{p_{\perp}}
\newcommand {\meanpT}	{\langle p_{\perp}\rangle}
\newcommand {\Rpt}	{R_{\meanpT}}
\newcommand {\RNch}	{R_{\mean{\Nch}}}

\newcommand {\mean}[1]	{\langle #1\rangle}

\newcommand {\rZr} {q_{\rm ZrZr}}
\newcommand {\rRu} {q_{\rm RuRu}}
\newcommand {\dQ} {\Delta Q}

\begin{document}
\title{Probing neutron skin and symmetry energy with relativistic isobar collisions}
%
%

\author{\firstname{Hao-jie} \lastname{Xu}\inst{1}\fnsep\thanks{\email{haojiexu@zjhu.edu.cn}} 
}

\institute{School of Science, Huzhou University, Huzhou, Zhejiang 313000, China
          }

\abstract{%
In these proceedings, we present the three proposed observables to probe the neutron skin and symmetry energy with relativistic isobar collisions, namely,
the isobar ratios of the produced hadron multiplicities ($\Nch$), the mean transverse momenta ($\mean{\pt}$), and the net charge multiplicities ($\Delta Q$).
Our findings suggest potentially significant improvement to neutron skin and symmetry energy determination over traditional low energy methods.
}
\maketitle
\section{Introduction}
\label{intro}
Recently, the STAR collaboration has published the first result of the chiral magnetic effect (CME) search in isobar \RuRu\ and \ZrZr\ collisions at nucleon-nucleon center-of-mass energy $\snn=200$ GeV~\cite{STAR:2021mii}. 
While no evidence is found for the CME, the experimental data  show  sizeable differences in  multiplicity and elliptic flow between isobar collisions,  indicating significant differences in their nuclear structures, such as the neutron skin thickness and nuclear deformation. 

The neutron skin thickness \rnp\ $\equiv\sqrt{\langle r_{n}^{2} \rangle}-\sqrt{\langle r_{p}^{2}\rangle}$, the root mean square difference between neutron  and proton distributions, is strongly correlated to the density slope parameter of the symmetry energy.
The symmetry energy and its density dependence are crucial to our understanding of the masses and drip 
lines of neutron-rich nuclei and the equation of state (EOS) of nuclear and neutron star matter.
The \rnp\ has traditionally been measured by low-energy electron and hadron scatterings off nuclei~\cite{Frois:1987hk,Tsang:2012se,Tarbert:2013jze}, and the symmetry energy slope parameters $L$ and $L_c$ are extracted at 
the nuclear saturation density $\rho_{0}$ and  critical density $\rho_{c}$, respectively~\cite{Chen:2005ti,Roca-Maza:2011qcr,Tsang:2012se,Horowitz:2014bja,Wang:2022cda}. Because of the inevitable uncertainties in modeling the strong interaction of the scattering processes in quantum chromodynamics (QCD)~\cite{Ray:1992fj}, large uncertainties on the $L$ and $L_c$ persist. 
The parity-violating electroweak scattering measurement on the \Pb\ by the Lead Radius Experiment (PREX-II), although void of QCD uncertainties, has relatively large statistical uncertainty, \rnp\ = $0.283\pm0.071$~fm~\cite{Abrahamyan:2012gp,Adhikari:2021phr}. It leads to 
$L=105\pm37$~MeV~\cite{Reed:2021nqk}, at a slight tension with  $L=75\pm25$~MeV~\cite{Centelles:2008vu} from traditional scattering experiments.

The sensitivity of relativistic isobar collisions to neutron skin, predicted~\cite{Xu:2017zcn,Li:2018oec} and confirmed  by the STAR data~\cite{STAR:2021mii}, offers a new opportunity to complement low-energy measurements with completely different systematics. 
Specifically, the ratios between isobar collisions of the produced hadron multiplicities ($\Nch$)~\cite{Li:2019kkh}, the mean transverse momenta ($\mean{\pt}$)~\cite{Xu:2021uar}, and the net charge multiplicities ($\Delta Q$)~\cite{Xu:2021qjw} are found to be sensitive to the neutron skin difference between the isobar nuclei. 
Measurements of those ratios can, in turn, offer an unconventional and perhaps more precise means to probe the neutron skin. 

\section{Symmetry energy and  density functional theory}
The symmetry energy encodes the energy related to neutron-proton asymmetry in the
nuclear matter EOS. 
It is conventionally defined in the binding energy per nucleon, approximately expressed as~\cite{Li:2008gp}
$E(\rho,\delta)=E_0(\rho)+E_{\mathrm{sym}}(\rho)\delta^2
+{\cal O}(\delta^4)$,
where $\rho=\rho_{n}+\rho_{p}$ is the nucleon number density
and $\delta=(\rho_{n}-\rho_{p})/\rho$ is the isospin asymmetry
with $\rho_{p}$ ($\rho_{n}$) denoting
the proton (neutron) density.
The symmetry energy can be obtained as
$E_{\rm sym}(\rho)=\left.\frac{1}{2}\frac{\partial^{2}E(\rho,\delta)}{\partial\delta^{2}}\right|_{\delta=0}.$
At a given reference density $\rho_r$, the $E_{\rm sym}(\rho)$ can be expanded in
$\chi_r=(\rho-\rho_r)/3\rho_{r}$ as
$E_{\rm sym}(\rho) = E_{\rm sym}(\rho_r) + L(\rho_r) \chi_r + \mathcal{O}(\chi_r^2)$,
where $L (\rho_r) = \left. 3\rho_r\frac{dE_{\rm sym}(\rho)}{d\rho}\right|_{\rho=\rho_r}$ 
is the density slope parameter~\cite{Li:2008gp}. 
Especially, for $\rho_r = \rho_0 \approx 0.16$~fm$^{-3}$ and  
$\rho_c = 0.11\rho_0/0.16 \approx 0.11$~fm$^{-3}$, 
one has $L \equiv L (\rho_0)$  and $L_{c}\equiv L (\rho_{c})$ which characterizes
the density dependence of the $E_{\rm sym}(\rho)$ around $\rho_0$ and $\rho_{c}$.
A strong constraint $L(\rho_c) = 47.3 \pm 7.8$~MeV is obtained from analyzing the data on 
the electric dipole polarizability  in $^{208}$Pb~\cite{Zhang:2014yfa}.
Generally, it is found that  the $L(\rho_c)$ displays a particularly
strong positive correlation with the \rnp\ of heavy nuclei.

We use two different nuclear energy density functionals to 
describe nuclear matter EOS and the properties of finite nuclei, namely,
the standard Skyrme-Hartree-Fock (SHF) model (see, e.g., Ref.~\cite{Chabanat:1997qh}) and the extended SHF (eSHF) model~\cite{Zhang:2015vaa}.
Compared to SHF,
the eSHF model contains additional momentum and density-dependent two-body forces
to simulate the momentum dependence of the three-body forces effectively~\cite{Zhang:2015vaa}.
Fitting to data using the strategy in Ref.~\cite{Zhou:2019omw}, we obtain
an interaction parameter set (denoted as Lc47) within eSHF by fixing $L(\rho_c) = 47.3$ MeV~\cite{Zhang:2014yfa} 
with $E_{\rm sym}(\rho_c) = 26.65$~MeV~\cite{Zhang:2013wna}.
We also construct two more interaction parameter sets (denoted as Lc20 and Lc70) with $L(\rho_c) = 20$~MeV and $70$~MeV, respectively,
keeping the same $E_{\rm sym}(\rho_c)$~\cite{Zhang:2013wna}, to explore the effects of the symmetry energy (and neutron skin) variations.
For the SHF calculations, we use the well-known interaction set SLy4~\cite{Chabanat:1997un,Wang:2016rqh}.
The four interaction parameter sets give similar proton rms radius $r_p$
for {\Zr } and {\Ru } which are experimentally well constrained,
but the neutron radius $r_n$ increases with $L(\rho_c)$ and $L$,
leading to a positive correlation between \rnp\ and $L(\rho_c)$ (and $L$) as expected.
Both DFT calculations predict a halo-type neutron skin thickness for \Zr, which is crucial to our understanding of non-trivial bump structures in the ratios of the multiplicity distributions and the elliptic flows in non-central isobar collisions~\cite{Xu:2017zcn,Li:2018oec,Xu:2021vpn,STAR:2021mii}

\section{Probes in isobar collisions}
\subsection{Total charge multiplicity $\Nch$}
The event multiplicity produced in heavy-ion collisions is sensitive to the density distributions of the colliding nuclei, and thus the $L$.
The absolute $\Nch$ values are subject to significant model dependence
because particle production in heavy ion collisions is generally hard to model precisely.
The shape of the $\Nch$ distribution is, on the other hand, more robust. It is determined by the interaction cross-section as a function of the impact parameter ($b$).
We use four typical, well developed, commonly used models for relativistic heavy ion collisions to simulate the particle production:
the \hijing\ (Heavy ion jet interaction generator, v1.411) 
model, the  \ampt\ (A Multi-Phase Transport, v1.26, v2.26) model 
with string fragmentation (v1.26) and string melting (v2.26), and the \urqmd\ (Ultra relativistic Quantum Molecular Dynamics, v3.4) 
model.
Except for the input nuclear density distributions, all parameters are set to default. 

The $\Nch$ distributions show  splittings with different $L_{c}$. We  use the relative $\mean{\Nch}$ difference between Ru+Ru and Zr+Zr,
	$\RNch = 2\frac{\mean{\Nch}_{\rm RuRu}-\mean{\Nch}_{\rm ZrZr}}{\mean{\Nch}_{\rm RuRu}+\mean{\Nch}_{\rm ZrZr}}$,
to quantify the splitting of the $\Nch$ tails. Experimental measurements of $\Nch$ are affected by tracking inefficiency,
usually multiplicity dependent~\cite{STAR:2008med}. While this effect  mostly cancels out in $\RNch$, it is better to use only central collisions, e.g. top $5\%$ centrality, where the tracking efficiency is constant to a good degree. 
\begin{figure} 
\begin{minipage}{0.55\textwidth}
    \centering
	\includegraphics[scale=0.32]{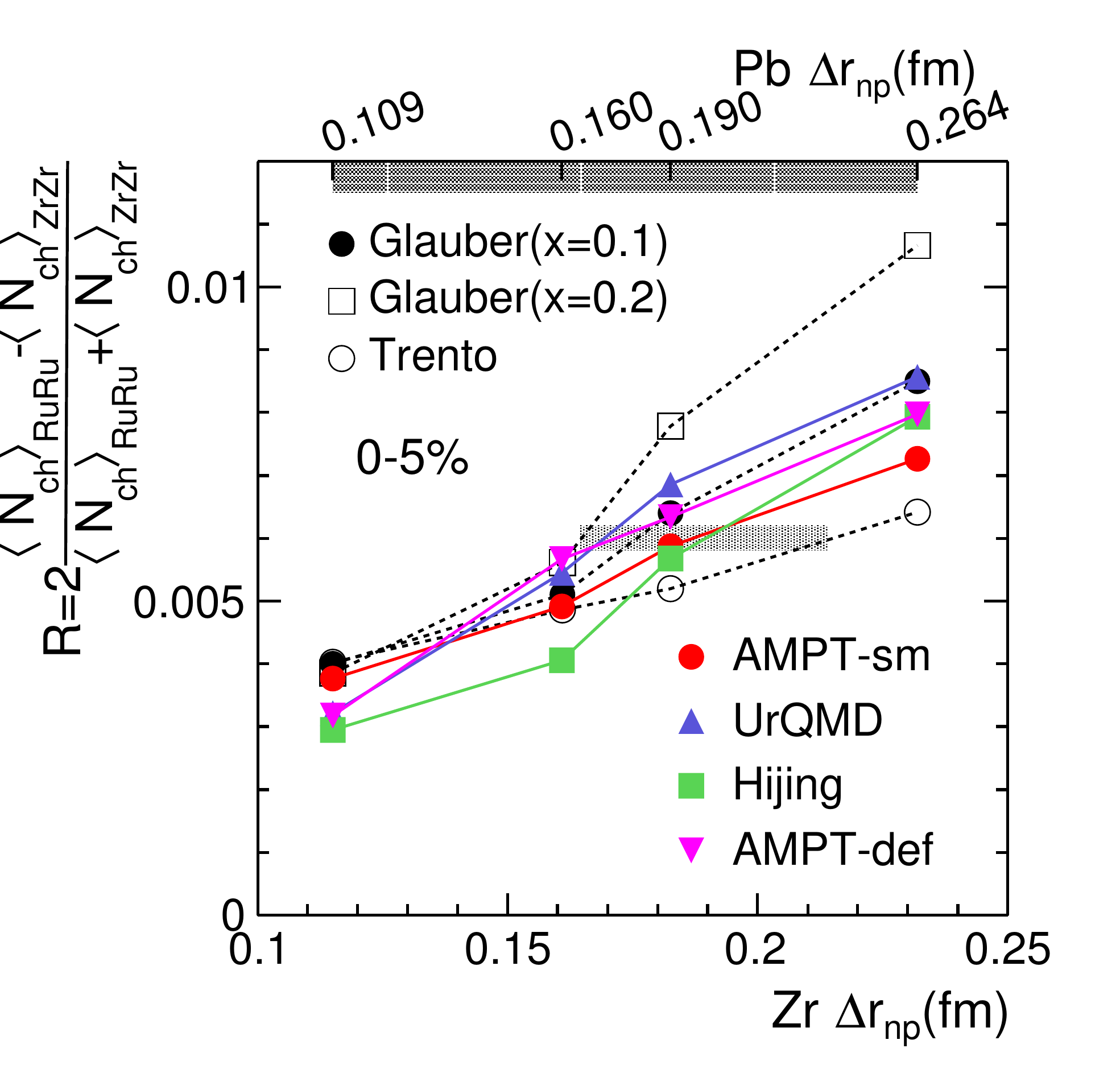}
\end{minipage}
\begin{minipage}{0.40\textwidth}
\vspace*{1cm}
	\caption{(Color online). The relative $\mean{\Nch}$ ratio $\RNch$ as a function of the Zr neutron skin thickness. The four sets of data points in order from left to right are from Lc20,  SLy4, Lc47, Lc70 densities. 
	The results from \amptsm, \urqmd, \hijing, and \amptdef\ are connected by solid lines.
		The results from Glauber and Trento models are connected by dashed lines. The figure is taken from Ref.~\cite{Li:2019kkh}.
	\label{fig:RatioNch}} 
\end{minipage}
\end{figure}

The $\RNch$ in each model must depend on how much the Ru and Zr nuclear density distributions differ, which can be characterized by \rnp\ of the Zr (or Ru) nucleus. 
We therefore plot in Fig.~\ref{fig:RatioNch} the $\RNch$ in the top $5\%$ centrality against \rnp\ of the Zr nucleus from the eSHF (SHF) calculations with Lc20, Lc47, and Lc70 (SLy4). It is found that $\RNch$ monotonically increases with \rnp. This is because, with increasing \rnp, the difference between Ru and Zr densities increases. This results in an increasing difference in $\Nch$ between Ru+Ru and Zr+Zr collisions. 
The value of $\RNch$ has a relatively weak model dependence, including  \hijing\  whose tail distribution is significantly narrower than the other models,
which can already be discriminated by data measurements. 
The intuitive geometrical models, namely, the Glauber and Trento models give a similar trend as the above dynamical models.
In addition, $\RNch$ is a relative measure between Ru+Ru and Zr+Zr collisions, and much of the experimental effects are canceled. 
We thus expect a precise constraint on symmetry energy with our proposed observable $\RNch$ in relativistic isobar collisions~\cite{Li:2019kkh}.

The $\RNch$ observable has been measured in top $2\%$ centrality by the STAR collaboration~\cite{Xu:2022ikx}. Based on the Monte Carlo Glauber model simulations with the density distributions obtained from eSHF calculations, the  symmetry energy slope parameter is extracted to be $L(\rho_{c})=53.8 \pm 1.7 ({\rm stat.}) \pm 7.8 ({\rm syst.})$~MeV, where the systematic  uncertainty is dominated by those on  nuclear deformations.

\subsection{Mean transverse momentum $\meanpT$}
Transverse momentum ($\pt$) generation in relativistic heavy ion collisions is sensitive to the initial geometry and the final-state bulk evolution.
In hydrodynamics, the $\meanpT$ values are sensitive to the medium bulk properties.
To investigate the effects of bulk properties, we calculate the $\meanpT$ using the Lc47 densities with three values of shear viscosity [$(\eta/s)_{\rm min}=0.04, 0.08$ and $0.16$] and with three values of bulk viscosity  [$(\zeta/s)_{\rm max}=0.025, 0.081$ and $0.1$] in the iEBE-VISHNU (an event-by-event (2+1)-dimensional viscous hydrodynamics, together with the hadron cascade model UrQMD) model simulations. 
The middle values are typical values used in hydrodynamic simulations~\cite{Bernhard:2019bmu}.
Our findings strongly indicate while the magnitude of $\meanpT$ depends on the bulk properties, the ratio
$	\Rpt \equiv \frac{\meanpT^{\rm Ru+Ru}} {\meanpT^{\rm Zr+Zr}}$
is insensitive to them and hence their  uncertainties. 

\begin{figure*} 
	\begin{centering}
      \includegraphics[scale=0.28]{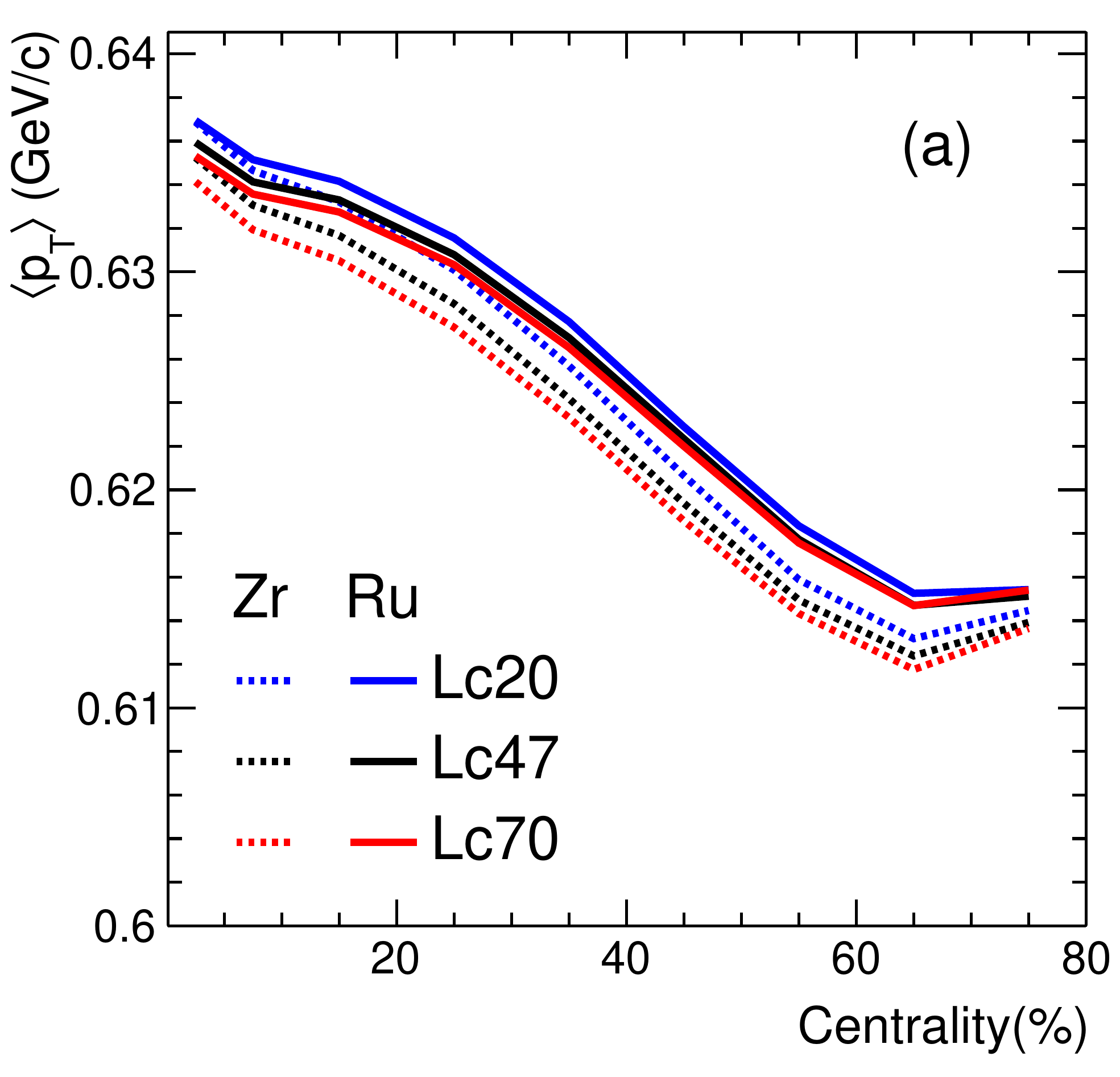}
	\includegraphics[scale=0.28]{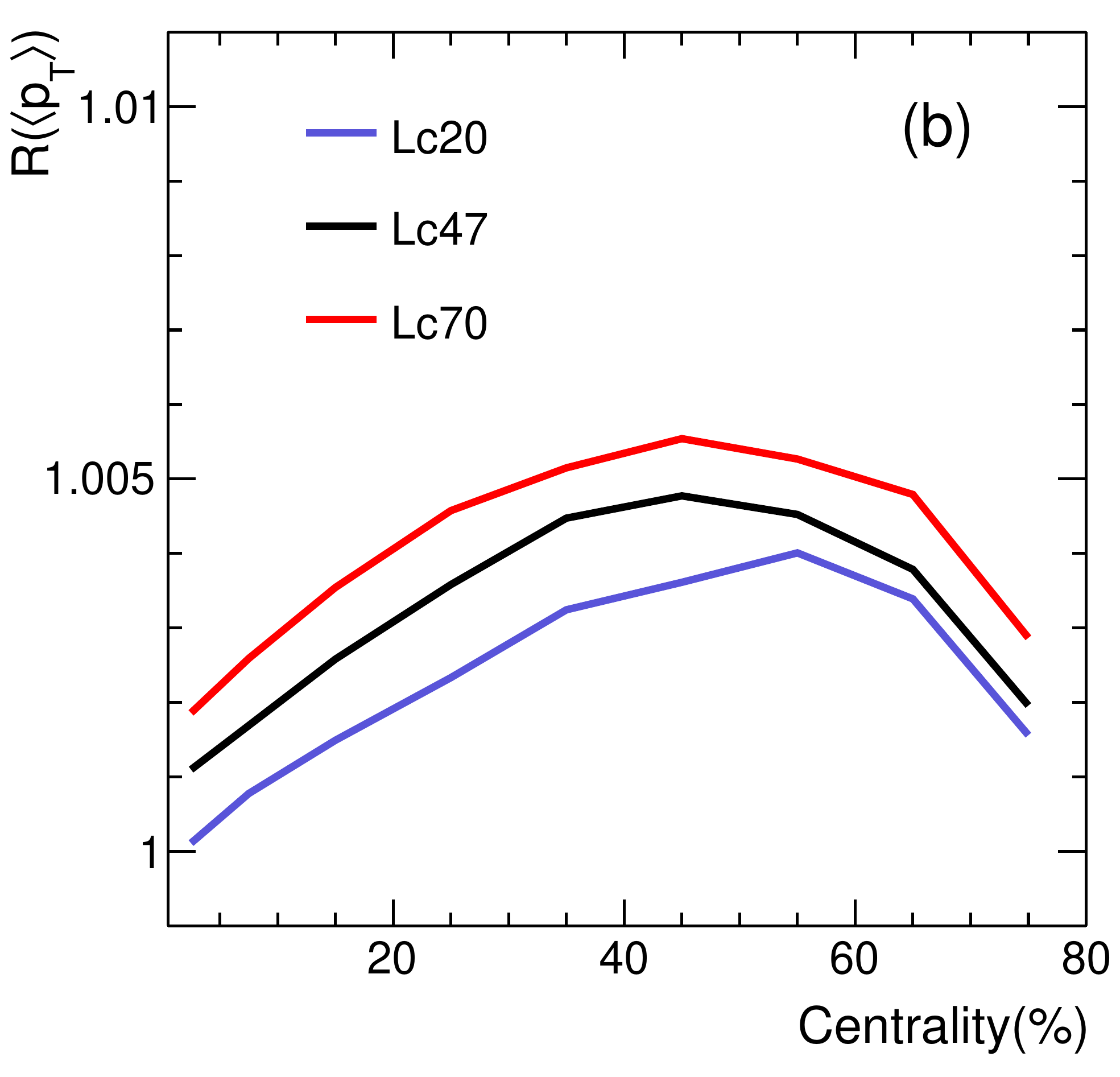}
      \caption{(Color online).
	  (a) The mean transverse momentum $\meanpT$ in Zr+Zr collisions, and (b) the \RuZr\ ratio $\Rpt$ as functions of centrality,  calculated by the iEBE-VISHNU model with Lc20, Lc47, and Lc70 spherical nuclear densities. Figures are taken from Ref.~\cite{Xu:2021uar}.
      \label{fig:meanpTA}} 
	\end{centering}
 \end{figure*}

Figure~\ref{fig:meanpTA}(a) presents the $\meanpT$ as functions of centrality in both Ru+Ru  and Zr+Zr collisions from the iEBE-VISHNU simulations with various DFT-calculated spherical densities for the isobars.
A larger $\Lc$ gives thicker neutron skin, and results in smaller $\meanpT$ at each centrality, as expected. 
On the other hand, the \RuZr\ ratio $\Rpt$, shown in Fig.~\ref{fig:meanpTA}(b), increases with $\Lc$. 
This is because the neutron skin effect is larger in \Zr\ than in \Ru\ and this effect increases with $\Lc$. 
The splittings shown in the figure indicate the sensitivity of $L_{c}$ on $\Rpt$, providing a novel method to constrain symmetry energy in relativistic isobar collisions~\cite{Xu:2021uar}.
The centrality dependence of $\Rpt$ is non-trivial and can reach as large as 0.5\% above unity.

This method has recently  been applied to the isobar data by the STAR collaboration~\cite{Xu:2022ikx}. Based on the $\Rpt$ values in the top $5\%$ centrality and state-of-the-art hydrodynamic simulations, the symmetry energy slope parameter has been extracted. The value is $L(\rho_{c})=56.8 \pm 0.4 ({\rm stat.}) \pm 10.4 ({\rm syst.})$ MeV, where the systematic uncertainty  is dominated by those on nuclear deformations. This result is consistent with that extracted from $\RNch$.

\subsection{Net charge $\dQ$}
We note that the final state $\Nch$ and $\mean{\pt}$ do not distinguish between initial neutron or proton participants.
Proton-proton ($pp$), proton-neutron ($pn$), and neutron-neutron ($nn$) collisions at relativistic energies produce practically the same $\mean{\Nch}$ and $\mean{\pt}$.
It is sensitive only to the overall nucleon density, and therefore indirectly sensitive to the neutron
density (and neutron skin) given that the proton density is well determined. But if nuclei had proton skin
instead of neutron skin, our study would yield the same result. 
Ultra-peripheral collisions, where the nuclei are only grazing each other, must have very different mixture of participant protons and neutrons, and therefore likely yield significantly different net-charge numbers ($\dQ$). 

\begin{figure} [!htb]
\begin{minipage}{0.55\textwidth}
	\centering
      \includegraphics[scale=0.30]{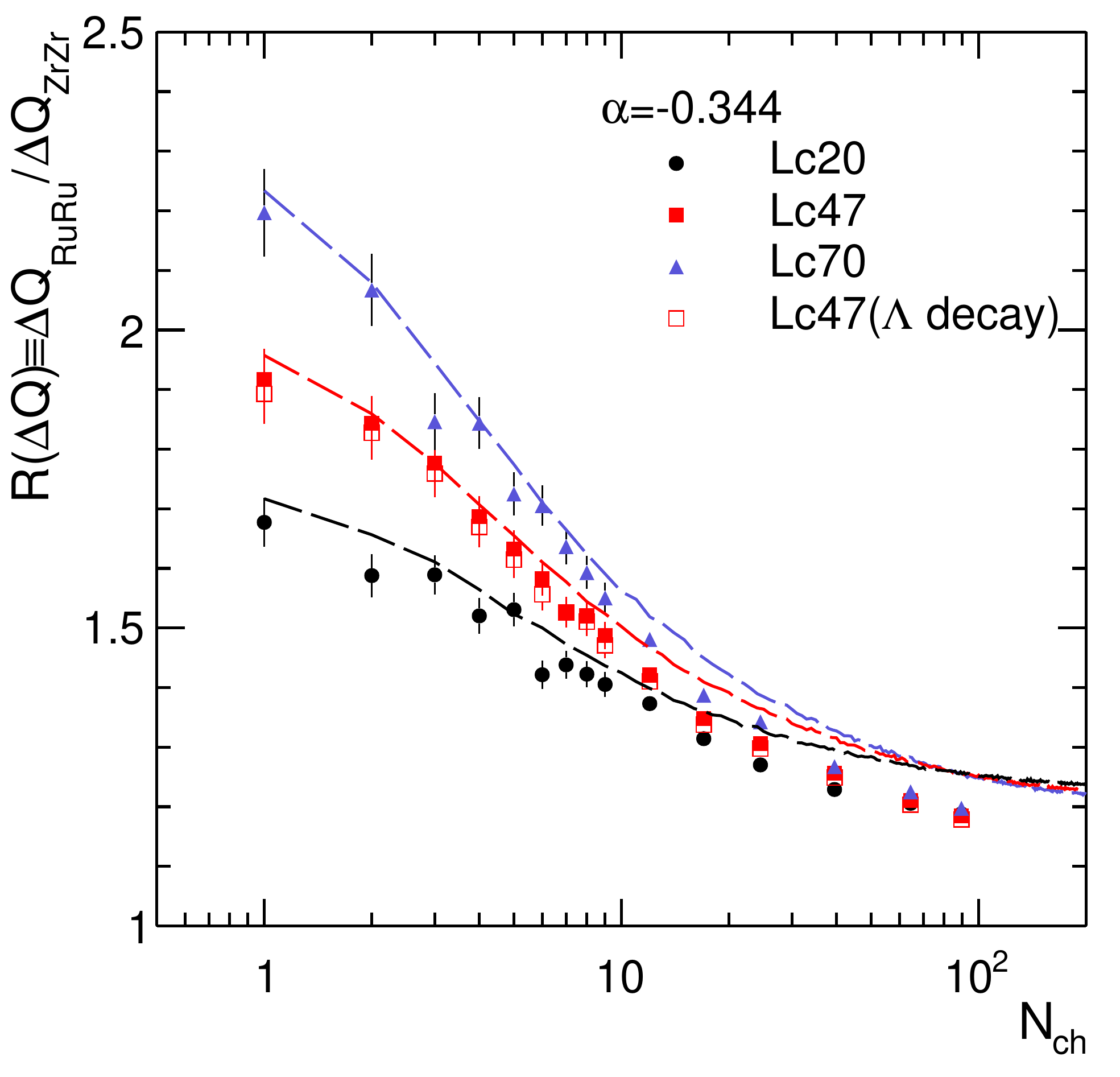}
\end{minipage}
\begin{minipage}{0.40\textwidth}
      \caption{(Color online).
$R_{\dQ}$ for $b\in[7,20]$~fm as function of $\Nch$ ($|\eta|<0.5$) simulated by \urqmd\ with DFT nuclear densities from eSHF (Lc20, Lc47, Lc70). 
The open red squares show a calculation for the Lc47 case including $\Lambda$-hyperon decays.
		The curves are from superimposition prediction with $\alpha=-0.344$ from \urqmd\ NN interactions and the $q_{AA}$ from Trento simulation. Figure is taken from Ref.~\cite{Xu:2021qjw}.
		\label{fig:RQUrQMD}}
\end{minipage}
\end{figure}

The $\dQ$ ratio in Ru+Ru over Zr+Zr collisions, under the superimposition assumption, is
      $R_{\dQ}\equiv\frac{\dQ_{\rm RuRu}}{\dQ_{\rm ZrZr}} 
      = \frac{\rRu + \alpha/(1-\alpha)}{\rZr + \alpha/(1-\alpha)}$, 
where $\alpha \equiv \Delta Q_{nn}/\Delta Q_{pp}$ is the $\dQ$ ratio in $nn$ to $pp$ interactions and $q_{\rm AA}$ is the fraction of protons among the participant
nucleons that can be obtained from Trento simulations. \pythia 
(version 8.240) gives $\alpha\simeq -0.352$ with acceptance cuts 
$|\eta|<1$ and $0.2<p_{T}<2$ GeV/$c$ excluding the (anti-)protons with $p_{T}<0.4$ GeV/$c$.
The overall $\rRu$ and $\rZr$ values for the whole nuclei are $44/96$ and $40/96$, respectively;
they would give $R_{\dQ}\simeq 1.267$. Of course, the simple superimposition assumption breaks down in non-peripheral collisions because of nuclear effects. However, the assumption should be good for grazing AA collisions.
The general idea to probe \rnp\ by $R_{\dQ}$ is that
a sizeable \rnp\ will make the $q_{AA}$ decrease dramatically with increasing impact parameter ($b$) in those grazing collisions. 
The \rnp\ of \Zr\ is significantly larger than that of \Ru, so the $R_{\dQ}$ ratio amplifies the \rnp\ sensitivity. The \rnp\ of both nuclei are controlled by the $L_c$ parameter, thus a measurement of $R_{\dQ}$ can determine its value.

We simulate $\sim$450 million events within $b\in[7,20]$~fm using \urqmd\ model. 
The same acceptance cuts have been applied as done in \pythia\ simulations. 
Figure~\ref{fig:RQUrQMD} shows $R_{\dQ}$ as a function of $\Nch$. 
Using $\alpha=-0.344$, the predicted curves from superimposition assumption are depicted in Fig.~\ref{fig:RQUrQMD}. The curves can fairly well describe the \urqmd\ data.
This indicates that the grazing collisions in \urqmd, with $\Nch\lesssim 10$, are indeed simple superimposition of NN interactions. This is not surprising as only a few nucleons participate in such a grazing AA collision, so any nuclear effect would be negligible.
At higher $\Nch$ the \urqmd\ data points deviate from the curves, presumably because those collisions are not simple NN superimpositions anymore.
It may also be viewed that the effective $\alpha$ in central AA collisions, because of nuclear effects, is very different from the one calculated using single NN interactions.
The splittings shown in the figure indicate the sensitivity of $L_{c}$ on $R_{\Delta Q}$, providing a novel method to constrain symmetry energy in relativistic isobar collisions~\cite{Xu:2021qjw}.

\section{Summary}
The isobar \RuRu\ and \ZrZr\ collisions at $\snn=200$ GeV provide novel means to probe the neutron skins of the isobar nuclei. 
The neutron skin thickness can be determined, with the help of DFT calculations, from the isobar ratios of
the produced hadron multiplicities ($\RNch$), the mean transverse momenta ($\Rpt$), and the net charge multiplicities ($R_{\Delta Q}$). 
Due to the rather weak dependence of  these ratios to the details of QCD, our proposed methods can be used to determine the density slope parameter of symmetry energy with a precision that may be comparable to or even exceed those achieved by traditional low-energy nuclear experiments. Our measurements  complement,  with  different systematics, those low-energy experiments in probing the symmetry energy. 
The preliminary results on the extracted  slope parameter $L(\rho_{c})$ from $\RNch$ and $\Rpt$ have been reported by the STAR collaboration~\cite{Xu:2022ikx}, with the values of  $53.8\pm1.7\pm7.8$~MeV and $56.8\pm0.4\pm10.4$~MeV, respectively.  
These values are consistent with world-wide data from traditional nuclear scattering experiments with comparable precision~\cite{Xu:2022ikx}.

This work is supported by National Natural Science Foundation of China (NSFC) under Grants No. 12275082, 12035006, 12075085, 11909059.

\end{document}